\documentclass[a4paper,11pt]{article}
\usepackage{pos}

\usepackage{multicol}
\usepackage{caption}

\title{Multi-Messenger studies with the Pierre~Auger~Observatory}

\manuallySeparateAuthors
\author*[a]{Lukas Zehrer}
\author[b,\dag]{ for the Pierre Auger Collaboration}

\affiliation[a]{University of Nova Gorica,\\
  Vipavska 13, 5000 Nova Gorica, Slovenia}

\affiliation[b]{Observatorio Pierre Auger,\\
Av. San Martiń Norte 304, 5613 Malargüe, Argentina}

\notes{\note{Full author list available at: \url{http://www.auger.org/archive/authors_2020_08.html}}}

\emailAdd{auger\_spokespersons@fnal.gov}

\abstract{Over the past decade the multi-messenger astrophysics has emerged as a distinct discipline, providing unique insights into the properties of high-energy phenomena in the Universe. The Pierre Auger Observatory, located in Malargüe, Argentina, is the world's largest cosmic ray detector sensitive to photons, neutrinos, and hadrons at ultra-high energies. Using its data, stringent limits on photon and neutrino fluxes at EeV energies have been obtained. The collaboration uses the excellent angular resolution and the neutrino identification capabilities of the Observatory for follow-up studies of events detected in gravitational waves or other messengers, through cooperation with global multi-messenger networks. We present a science motivation together with an overview of the multi-messenger capabilities and results of the Pierre Auger Observatory.}

\FullConference{%
  40th International Conference on High Energy physics - ICHEP2020\\
  July 28 - August 6, 2020\\
  Prague, Czech Republic (virtual meeting)
}

\begin{document}
\maketitle

\section{Introduction}

The earliest known messengers taking part in the strong interaction at the very high energies (VHE, 100~GeV to 100~TeV) are called cosmic rays (CRs), which were detected in 1912 by Victor Hess with his balloon experiment. For the messengers of the electromagnetic (EM) force at these energies the first photons $\gamma$ were reported in \cite{VHEgamma}, whilst the messengers that undergo the weak interaction, neutrinos $\nu$, were first reported in \cite{firstNU1,firstNU2}. In 2016 the first signals from the merging of two black hole (BH) systems, GW150914 \cite{AbbottGW}, were reported. The detection of gravitational waves (GWs) from the merging of binary systems has not only marked the birth of gravitational wave astronomy, but enabled to combine the messenger particles of all four of nature's fundamental forces and thereby completed multi-messenger (MM) astrophysics.

The Pierre Auger Observatory \cite{auger2} was designed with a main purpose to study the properties of CRs with unprecedented accuracy up to the highest energies. As it is shown in the presented results and plots in the following sections, the Pierre Auger collaboration is also able to investigate other messengers, like neutrinos, photons and galactic neutrons, is complementary to other observatories for those messengers, and has the capabilities to set the most stringent limits in certain ranges of energy.

\section{Neutrinos}


Concerning their physical properties, neutrinos are the ideal messenger to study the high energy universe, and searches at ultra-high energy (UHE; >~0.1~EeV) for these particles have been undertaken several times by the Pierre Auger Observatory \cite{neutrinos1,neutrinos2}.

The search for neutrino candidate events is performed separately for Earth-skimming (ES, zenith angle range $90^{\circ} \leq \theta \leq 95^{\circ} $) and down-going low (DGL, $60^{\circ} \leq \theta \leq 75^{\circ}$) and high (DGH, $75^{\circ} \leq \theta \leq 90^{\circ}$) neutrino-induced showers \cite{DG1,DG2}. 
The ES is the most effective channel for neutrino detection, since the target mass provided by the Earth is large compared to the atmosphere and also because the threshold energy is low.

The discrimination from other primary particles is based on the fact that neutrinos can potentially interact at any point in the atmosphere for any zenith angle, which means they are, in contrast to all other particles, able to interact very deep in the atmosphere. This leads to different signals in the Pierre Auger Surface Detector (SD) array, than in the case of UHECR-induced extensive air showers (EASs) \cite{neutrinos1}.


\subsection{Limits on diffuse fluxes and point-like sources}

\begin{figure}
\noindent
	\begin{tabular}{ll}
	a) & b) \\
	\includegraphics[height=54mm]{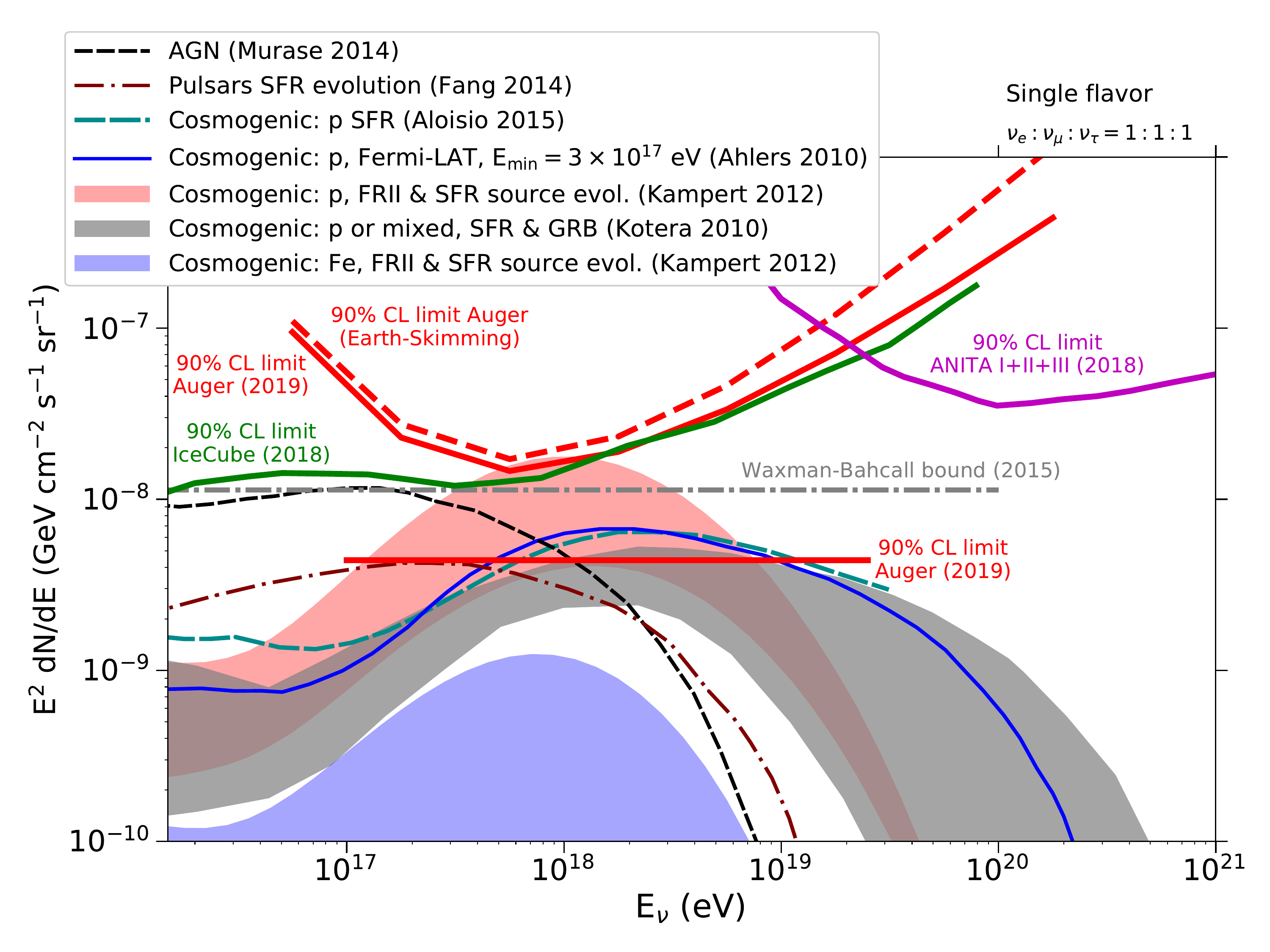} &
	\includegraphics[height=53mm]{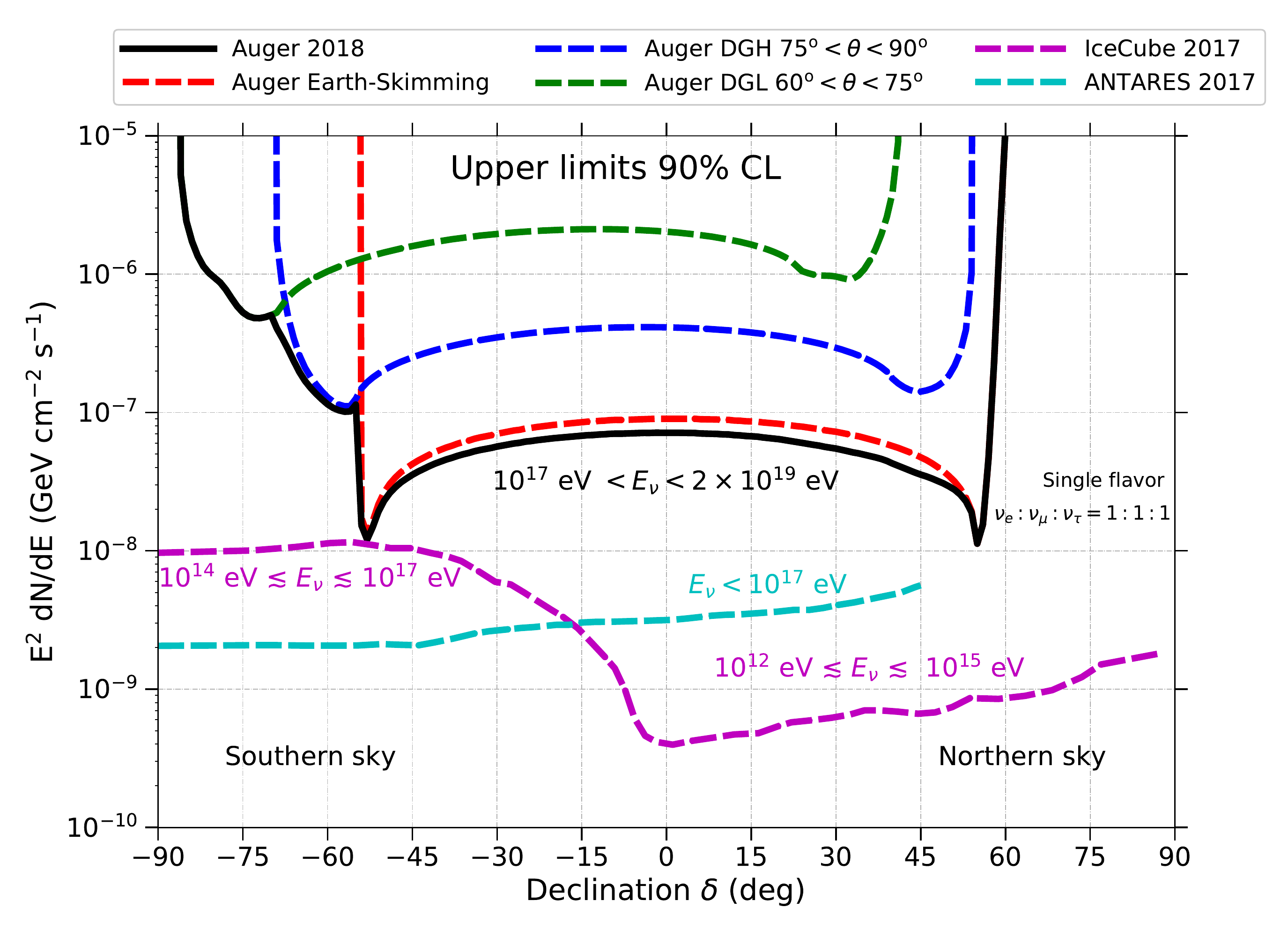} \\
	\end{tabular}
	\caption{\textbf{a)} Upper limit on diffuse fluxes, compared to the differential limits obtained by other experiments and astrophysical and cosmogenical neutrino models, showing in red the integral and differential upper limit for the normalisation constant assuming a $E^{-2}$ energy flux for a single flavour for all the three neutrino channels. The dashed red line is the Earth skimming channel alone, which contributes the strongest to the limit \cite{neutrinos2}. \textbf{b)} Auger 90\% CL upper limit on the fluxes of point-like sources as a function of equatorial declination obtained from the non-observation of ES and DGH neutrino candidates. Note the different energy ranges where the limits of each observatory apply. See \cite{neutrinos1} and references therein.}
	\label{fig:neutrinoLimits}
\end{figure}

The implications of the non-observation of UHE neutrinos are upper limits on diffuse fluxes, shown in figure \ref{fig:neutrinoLimits} a), and upper limits on flux from point-like sources of neutrinos, as can be seen in fig.~\ref{fig:neutrinoLimits} b).

The limit on diffuse fluxes is moving toward a physics scenario, which disfavours proton primary and strong evolution of the sources with redshift. In turn, a scenario of a mixed or heavy composition with weak evolution becomes more likely, which implies a suppressed neutrino flux. It can also be seen, that the differential upper limit on the diffuse neutrino flux of the Pierre Auger Observatory is comparable at EeV energies to the limits given from IceCube.

Neutrinos from point-like sources across the sky, with peak sensitivities at declination around $-53^{\circ}$ and $+55^{\circ}$, can be detected with the SD array. Since no neutrinos have been identified, upper limits on the neutrino flux from point-like steady sources have been calculated as a function of declination (see fig.~\ref{fig:neutrinoLimits} b)). In the plot, upper limits from the Pierre Auger Observatory for a single flavour point-like flux of UHE neutrinos is shown as a function of declination  together with IceCube and ANTARES upper limits at lower energies.

\subsection{UHE Neutrino follow-up of GW events}

\begin{figure}
\noindent
	\begin{tabular}{ll}
	a) & b) \\
	\includegraphics[height=48mm]{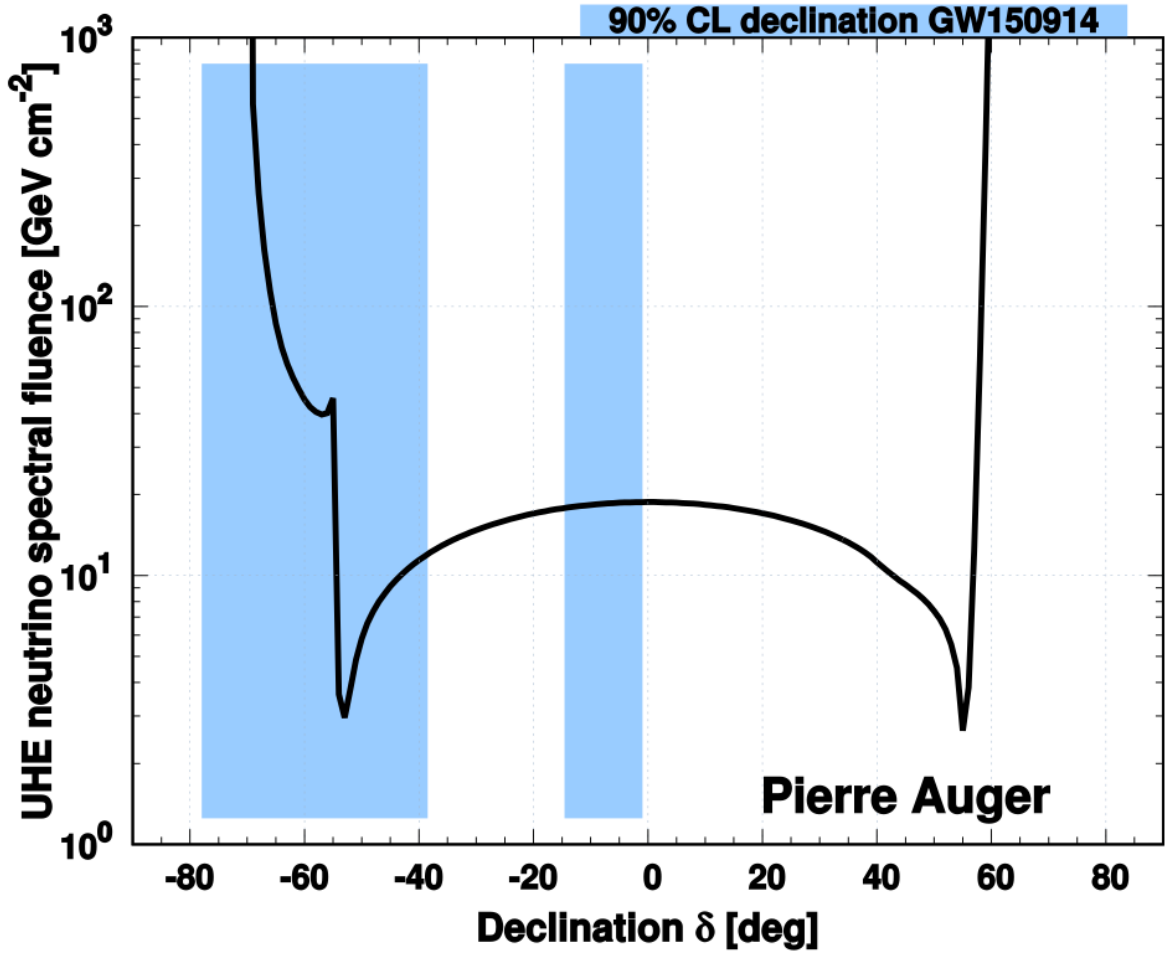} &
	\includegraphics[height=48mm]{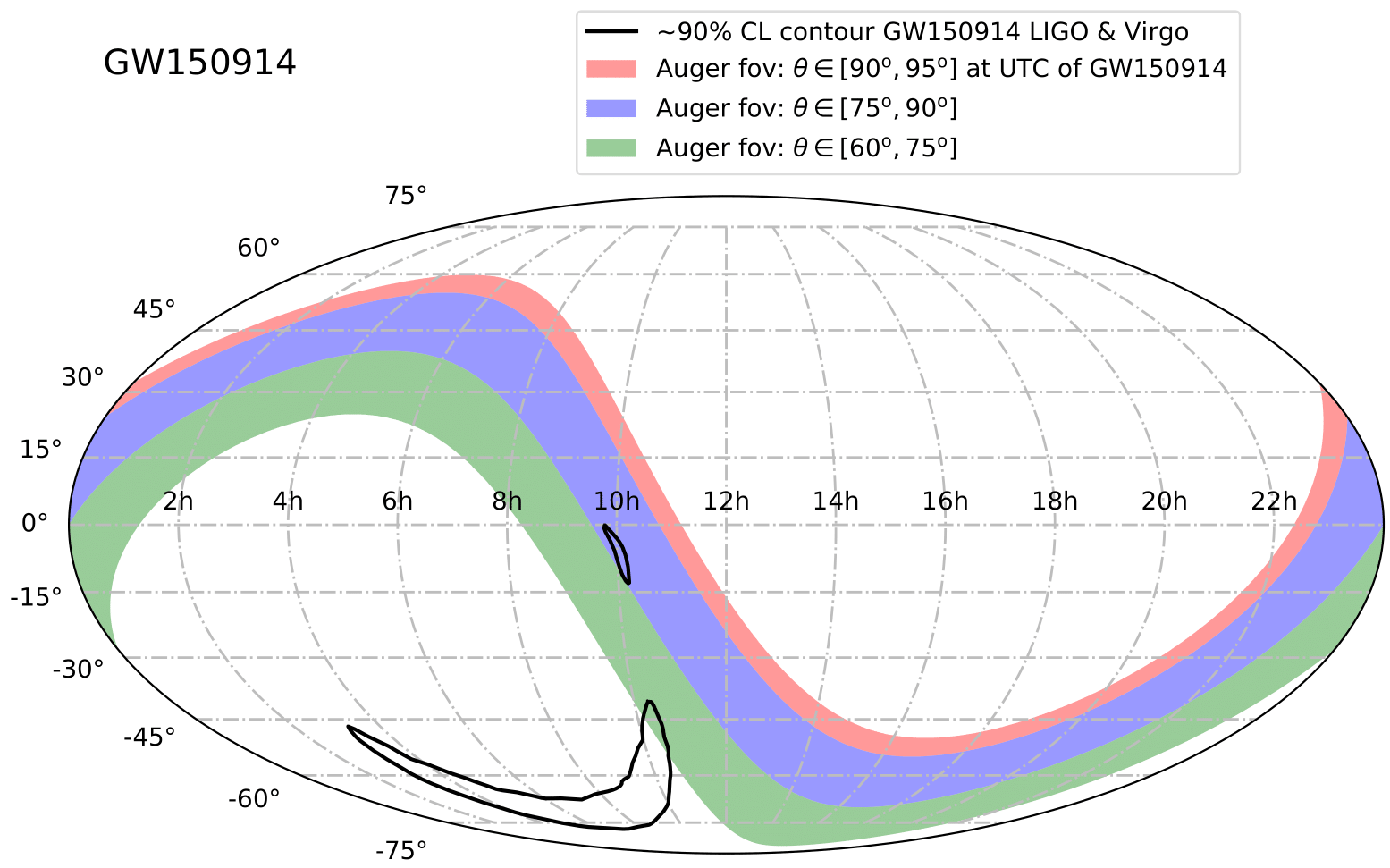} \\
	\end{tabular}
	\caption{\textbf{a)} In black the upper limit to the neutrino spectral fluence in the 100~PeV to 25~EeV range as a function of the equatorial source declination $\delta$, for the detection of the BH merger GW150914 is shown. The blue band is the 90\% C.L. sky localization of the reconstructed source \cite{GWfollow1}. \textbf{b)} Field of view of the Observatory at the instant of detection of the BH coalescence event GW150914 by LIGO, with band limits to separate the neutrino search into ES, DGH, and DGL channels, together with the 90\% C.L. region of the reconstructed position of the BH merger as obtained by LIGO observations (in black contours) \cite{multiAuger}.}
	\label{fig:GW150914}
\end{figure}

The Auger default neutrino search was applied around ±500~s and until 1 day after the merger inside a 90\% C.L. of most likely source location in the sky. The results for GW150914, the first GW event from a compact binary coalescence ever detected, are shown in fig.~\ref{fig:GW150914} a). 
The optimal declination positions $\delta \simeq -53^{\circ}$ and $\delta \simeq 55^{\circ}$, near the extremes in declination of the ES band, can be seen.

The field of view of the observatory at any given instant for each of the ES, DGH and DGL channels is limited to the bands corresponding to the zenith angle range of the channel, as displayed in fig.~\ref{fig:GW150914} b) \cite{multiAuger}.

\begin{figure}
\noindent
	\begin{tabular}{ll}
	a) & b) \\
	\includegraphics[height=55mm]{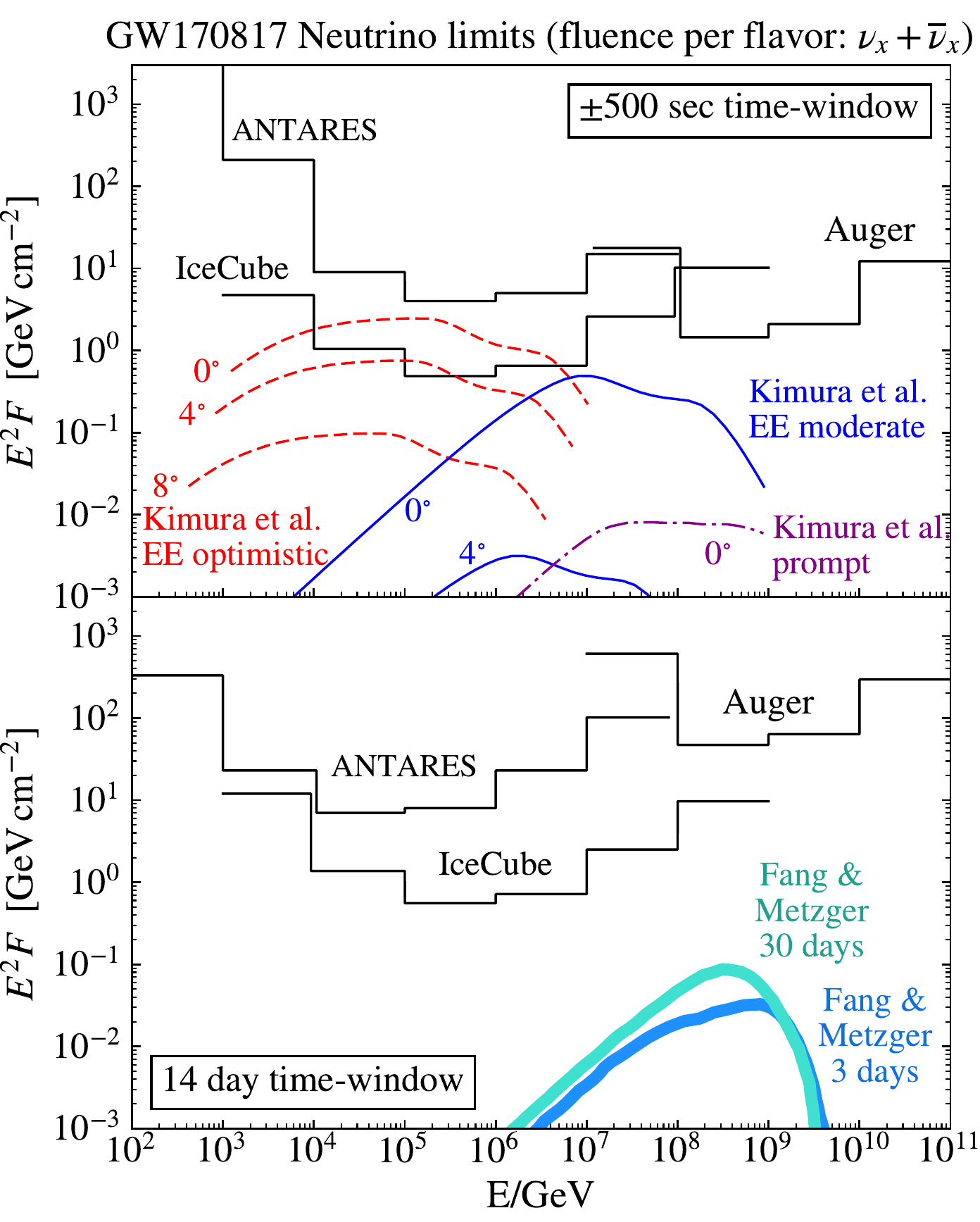} &
	\includegraphics[height=49mm]{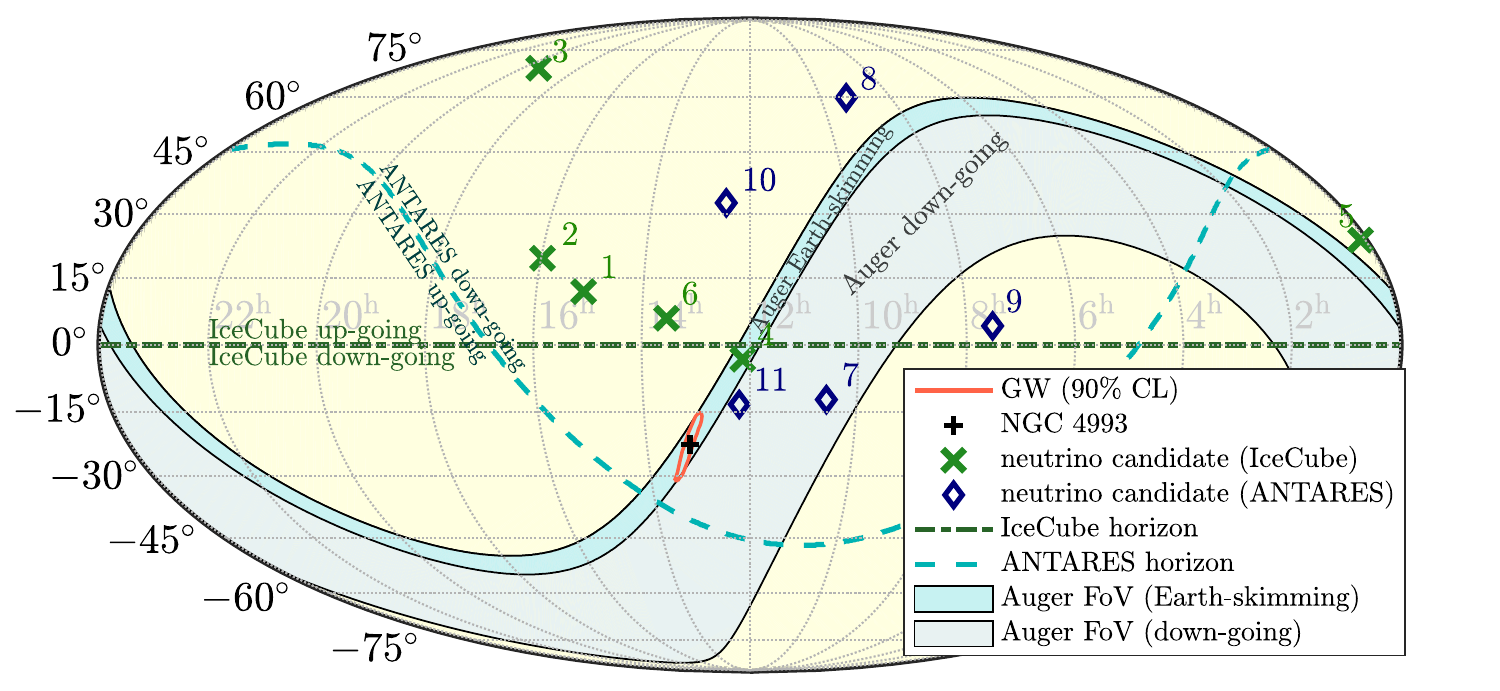} \\
	\end{tabular}
	\caption{\textbf{a)} \textbf{Top}: 90\% C.L. upper limits on the neutrino spectral fluence during a ±~500~s window as a function of energy in black angular lines. \textbf{Bottom}: Idem for a 14-day window following the GW trigger. The coloured smooth lines represent a model of UHE neutrino spectral fluences \cite{GW170817}. \textbf{b)} The sky location in equatorial coordinates of the GW170817 event (red contour) and its progenitor NGC4993, altogether with the Auger sensitive sky areas for neutrino detection at the time of the merger \cite{GW170817}.}
	\label{fig:GW170817}
\end{figure}

As can be seen in fig.~\ref{fig:GW170817} a), Auger provides the most stringent upper limit to the neutrino fluence at 90\% C.L. in the range from 100~PeV to 25~EeV to the GW event 170817, complementary to IceCube and ANTARES \cite{GW170817}. This GW event was produced by the merging of two neutron stars (NSs). In a time interval of $\pm$500~s around the NS merger its location was in the optimal Auger ES analysis window (93.3$^{\circ} < \theta < 90.4^{\circ}$).

In fig.~\ref{fig:GW170817}~b) the Auger field of view for ES and DG neutrino channels is shown for the event GW170817, altogether with the merger progenitor NGC4993 sky location, and the sky positions of neutrino candidate events that have been detected within 500~s of the merger.

\section{UHE Photons}

Photons at UHEs are considered the key to reveal the origin of the most energetic cosmic rays. A flux of UHE photons is expected to be produced from the interaction of nuclei, that can propagate over several Mpc, with the cosmic microwave background (CMB). Photons can also help to distinguish between scenarios of CR production via astrophysical particle acceleration ("bottom-up") vs. exotic models ("top-down").

Primary photons produce mostly EM showers with minor photo-nuclear or muon-pair production. With respect to protons and nuclei, EASs induced by photons have a deeper shower development and a smaller muon content. As a consequence, they differ also in other observable characteristics such as a steeper lateral distribution function, a smaller number of triggered stations, a slower rise of the signal in the SD stations on the ground, and a broader time front.

\subsection{Limit on the diffuse UHE photon flux}

With the Pierre Auger Observatory stringent limits are set on the diffuse flux of UHE photons. Above 10~EeV, SD data collected between 2004 and mid-2018 with an exposure of 40,000~km$^2$~sr~yr are used, and below 1~EeV unprecedented separation power between primary photons and hadrons can be achieved by combining observables from low-energy enhancements of the Pierre Auger Observatory. Altogether, the range of photon searches at the Pierre Auger Observatory extends to about three decades in energy.

\begin{figure}
	\centering
	\includegraphics[scale=0.19]{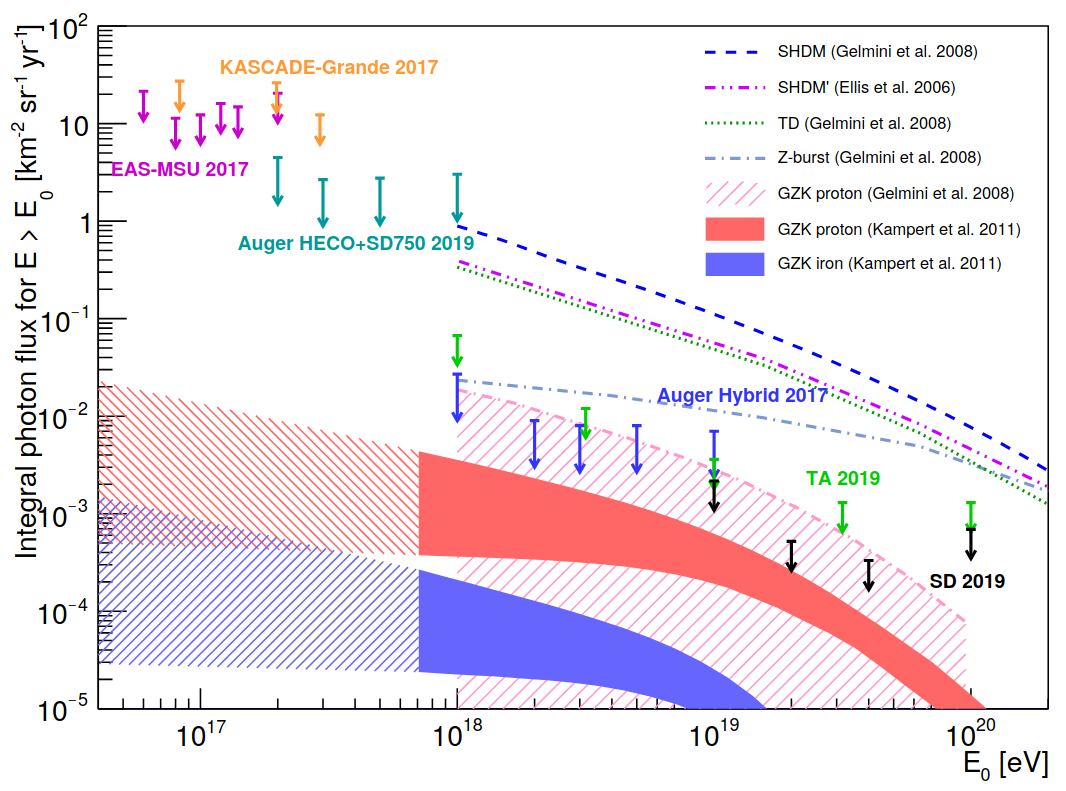}
	\caption{Limits on the photon flux at 95\% C.L. of the Pierre Auger Observatory, compared to model predictions and to the limits by other experiments \cite{photons}.}
	\label{fig:photons}
\end{figure}

Various variables with separation power are combined in a multivariate analysis (MVA) using a Boosted Decision Tree (BDT). Above 10~EeV, in a zenith angle range of 30$^{\circ} < \theta < 60^{\circ}$, 11 events were found to be above the threshold, two of them at an energy above 20~EeV, but conservatively an upper limit on the photon flux at 95\% C.L. was determined. The resulting upper limits on the integral photon flux for the thresholds of 0.2, 0.3, 0.5, and 1~EeV at 95\% C.L. can be seen in fig.~\ref{fig:photons} \cite{photons}. The Pierre Auger Observatory is thus the most sensitive air-shower detector for primary photons with energies above $\sim$0.2~EeV.

\section{Galactic Neutrons}

Neutron-induced EASs cannot be distinguished from the ones induced by charged CRs  on the basis of the shower development. Nevertheless, due to their non-deflected paths, it is in principle possible to determine sources of neutrons by identifying an excess from given directions, or by exploiting potential correlations in time and direction. Since the mean travel distance for relativistic neutrons with energies $E_n$ of a few EeV is 9.2~kpc $\cdot E_n/\mathrm{EeV}$, the distance from Earth to the Galactic center is about 8.3~kpc, and the radius of the Galaxy is approximately 15~kpc, neutrons at these UHEs can reach Earth from the entire Galaxy but not from much further away \cite{multiAuger}.

The choice of photon sources as probable neutron sources is motivated by the fact that both messengers are produced in photo-hadronic interaction scenarios. No significant excess of a neutron flux has been found in the searches from any class of candidate sources, however, strong limits at 95\% C.L. upper limits on the energy flux in neutrons have been deduced, thereby setting strong constraints on UHE proton production in our galaxy \cite{multiAuger}.

\section{Outlook}

We presented here an overview of the multi-messenger capabilities and results of the Pierre Auger Observatory.

By virtue of these achievements, Auger is contributing to global networks of observatories, such as the Deeper, Wider, Faster program (DWF, \href{http://www.dwfprogram.altervista.org}{www.dwfprogram.altervista.org}) and the Astrophysical Multi-messenger Observatory Network (AMON, \href{https://www.amon.psu.edu}{www.amon.psu.edu}), with possibilities to follow up and/or
to send a trigger for potentially interesting astrophysical MM events. 

The Auger capacity as a MM observatory is expected to further increase after the completion of the detector upgrade (AugerPrime \cite{upgrade1}) currently ongoing.

\end{document}